\documentclass[twocolumn,amsmath,amssymb,prl,superscriptaddress,longbibliography,nobalancelastpage]{revtex4-1}

\usepackage{titlesec}
\usepackage[colorlinks,bookmarks=false,citecolor=blue,linkcolor=blue,urlcolor=blue]{hyperref}
\usepackage{epsfig}
\usepackage{color}
\usepackage{mathtools}
\usepackage{subfigure}
\usepackage{multirow}
\usepackage{amsmath}
\usepackage{amsmath}
\usepackage{rotating}
\usepackage{lipsum}
\usepackage{graphicx} 
\usepackage{dcolumn}  
\usepackage{mwe}
\usepackage{mathtools}
\usepackage{multirow}
\usepackage{physics}
\newcommand\numberthis{\addtocounter{equation}{1}\tag{\theequation}}
\usepackage{braket}
\input{insbox}
\usepackage[titletoc,title]{appendix}
\usepackage[graphicx]{realboxes}
\usepackage{threeparttable}

\usepackage[utf8]{inputenc}
\usepackage[T1]{fontenc}

\newcommand{\KCF}{KCuF$_3$ }

\begin{document}

    \title{{Unravelling the orbital physics in a canonical orbital system \KCF}}  
    
    \author{Jiemin Li}
    \thanks{Present Address: National Synchrotron Light Source II, Brookhaven National Laboratory, Upton, NY, USA}
    \affiliation{Diamond Light Source, Harwell Campus, Didcot OX11 0DE, United Kingdom}
    \affiliation{Beijing National Laboratory for Condensed Matter Physics and Institute of Physics, Chinese Academy of Sciences, Beijing 100190, China}
    \author{Lei Xu}
    \thanks{Present Address: Theoretical Division, Los Alamos National Laboratory, Los Alamos, New Mexico 87544, USA}
    \affiliation{Institute for Theoretical Solid State Physics, IFW Dresden, Helmholtzstrasse 20, D-01069 Dresden, Germany}
    
    \author{Mirian Garcia-Fernandez}
    \affiliation{Diamond Light Source, Harwell Campus, Didcot OX11 0DE, United Kingdom}
    \author{Abhishek Nag}
    \affiliation{Diamond Light Source, Harwell Campus, Didcot OX11 0DE, United Kingdom}
    \author{H. C. Robarts}
   \affiliation{Diamond Light Source, Harwell Campus, Didcot OX11 0DE, United Kingdom}
    \affiliation{H. H. Wills Physics Laboratory, University of Bristol, Bristol BS8 1TL, United Kingdom}
    \author{A. C. Walters}
    \affiliation{Diamond Light Source, Harwell Campus, Didcot OX11 0DE, United Kingdom}
    \author{X. Liu}
    \affiliation{School of Physical Science and Technology, ShanghaiTech University, Shanghai 201210, China}
    \author{Jianshi Zhou}
    \affiliation{The Materials Science and Engineering Program, Mechanical Engineering, University of Texas at Austin, Austin, Texas 78712, USA}
    \author{Krzysztof Wohlfeld}
    \affiliation{Institute of Theoretical Physics, Faculty of Physics, University of Warsaw, Pasteura 5, PL-02093 Warsaw, Poland}
    \author{Jeroen van den Brink}
    \affiliation{Institute for Theoretical Solid State Physics, IFW Dresden, Helmholtzstrasse 20, D-01069 Dresden, Germany}
    \affiliation{Institut f{\"u}r Theoretische Physik and W{\"u}rzburg-Dresden Cluster of Excellence ct.qmat, Technische Universit{\"a}t Dresden, 01062 Dresden, Germany}
    \author{Hong Ding}
    \affiliation{Beijing National Laboratory for Condensed Matter Physics and Institute of Physics, Chinese Academy of Sciences, Beijing 100190, China}
    \author{Ke-Jin Zhou}
    \email{Corresponding author: \\ kejin.zhou@diamond.ac.uk}
    \affiliation{Diamond Light Source, Harwell Campus, Didcot OX11 0DE, United Kingdom}
    
        \date{\today}

    \begin{abstract}
        We explore the existence of the collective orbital excitations, orbiton, in the canonical orbital system‚ KCuF$_3$ using the Cu $L_3$-edge resonant inelastic X-ray scattering. We show that the non-dispersive high-energy peaks result from the Cu$^{2+}$ $dd$ orbital excitations. These high-energy modes display good agreement with the {\it ab-initio} quantum chemistry calculation indicating that the $dd$ excitations are highly localized. At the same time, the low-energy excitations present clear dispersion. They match extremely well with the two-spinon continuum following the comparison with Mueller Ansatz calculations. The localized $dd$ excitations and the observation of the strongly dispersive magnetic excitations suggest that orbiton dispersion is below the resolution detection limit. Our results can reconcile with the strong {\it local} Jahn-Teller effect in KCuF$_3$, which predominantly drives orbital ordering. 
    \end{abstract}

    \maketitle
    
\textit{Introduction.}---Similar to the spin or charge ordering, the electron orbital can form long-range ordering in strongly correlated materials \cite{Tokura2000}. For example, the colossal magnetoresistive manganite presents unusual transport properties that appear to be connected to its spin and orbital order (OO) coupling \cite{Ling2000}. In vanadates, the OO is known to be related to the multiple temperature-induced magnetization  \cite{Ren1998}. As one of few pseudocubic perovskite systems, \KCF has been reported to form a long-range OO at a temperature of about 800 K and undergo a three-dimensional (3D) antiferromagnetic (AFM) ordering below $T_N$ of 38 K \cite{Paolasini2002, Hutchings1969,Tsukuda1972,Caciuffo2002,Satija1980}. Along with manganites, \KCF is generally considered to be the most prototypical orbital ordered system.
        
The signature of the long-range spin order is the collective spin wave due to the superexchange interaction. Similarly, Kugel and Khomskii proposed the spin-orbital superexchange coupling from which a collective orbital excitation – orbiton – should in principle be concomitant with the presence of the OO in a correlated system \cite{Kugel1982}. One of the best studied 3D OO systems is LaMnO$_3$ in which the dispersive orbiton has been theoretically predicted \cite{Brink1999}. Raman scattering reported the orbitons in LaMnO$_3$, however, its existence has not been verified by other experiments, possibly due to the complex multiplet structure of the manganese ions \cite{Saitoh2001,Grueninger2002}. Thus, it is instructive to look at the other prototypical orbital system, \KCF, which is actually far simpler. Here the OO is generally accepted to be largely driven by the Jahn-Teller (JT) effect~\cite{Pavarini2008}. The search for orbitons by the nonresonant inelastic x-ray scattering (IXS) did not reveal any evidences in the energy range up to 120 meV. It has been argued, though, that the orbitons may exist at a much higher energy range ~\cite{Tanaka2004}. High-energy excitations have been studied by Cu $K$-edge resonant inelastic x-ray scattering (RIXS), but these studies contained no discussions of the existence of orbitons ~\cite{Ishii2011}. Interestingly, a large orbiton dispersion has been observed in various spin-orbital entangled cuprates and titanates by resonant inelastic soft x-ray scattering at the $L$ edges of transition metals owing to x-ray's sensitivity to spin and orbital excitations \cite{Schlappa2012, Bisogni2015, Fumagalli2020, Ulrich2008, Wohlfeld2011, Wohlfeld2013}.

In this Letter, we employ high-energy-resolution RIXS to explore the existence of orbitons in \KCF at the Cu $L_3$ edge. $L$-edge RIXS is a well-established method for directly probing the $dd$ orbital excitations and the collective orbital and magnetic excitations in transition metal oxides \cite{Schlappa2012, Bisogni2015, Fumagalli2020, Ulrich2008, Sala2011, Braicovich2010, Luuk2011}. It is therefore ideal to apply $L$-edge RIXS to shed light on the orbital physics in KCuF$_3$ in both the low- and high-energy regimes. At high energy, $dd$ excitations from nondegenerated Cu$^{2+}$ 3$d$ orbitals are resolved. They are nondispersive in the reciprocal space but demonstrate remarkable evolution in intensity. Using an \textit{ab initio} quantum chemistry calculation based on a single CuF$_6$ cluster, we reproduced $dd$ excitations successfully, indicating that the local crystal-field splitting induced by a JT distortion dominates the high-energy $dd$ excitations rather than the collective orbitons. At the low-energy range, dispersive excitations are clearly resolved. Through Muller ansatz calculations, we conclude the dispersive excitations in the low-energy range are dominated by the two-spinon continuum. Our results suggest that orbitons, if they exist, may appear at a much lower energy scale than theoretically expected.

        \begin{figure}
            \centering
            \resizebox{8.6cm}{!}
            {\includegraphics{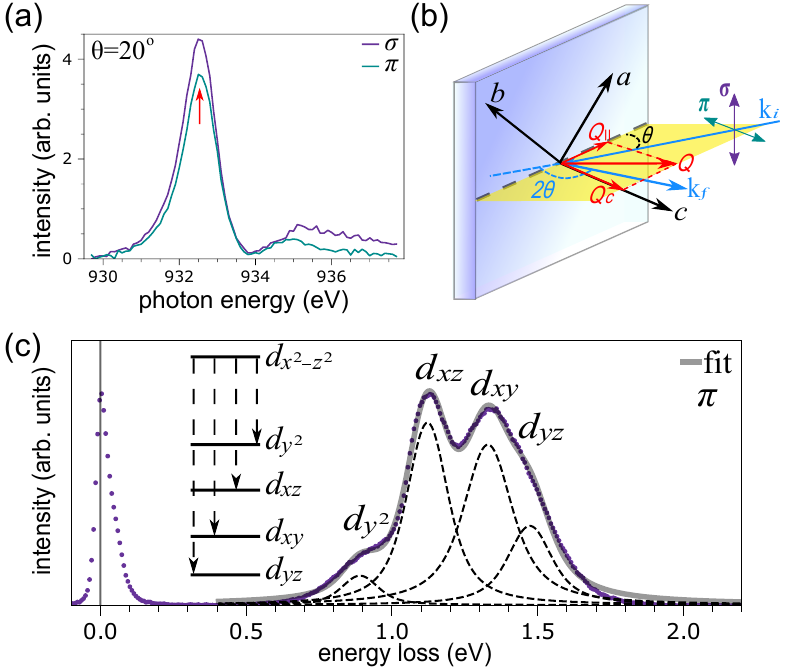}}
            \caption{Scattering condition and overview of RIXS spectra. (a) Cu $L_3$-XAS spectra of \KCF collected with the partial fluorescence yield. (b) A sketch of the experimental geometry. Light blue arrows represent the incident ($k_i$) and scattered ($k_f$)  x-rays, while the double arrows (green: $\pi$, purple: $\sigma$) are for the polarizations of incident x-rays. Red arrows indicate the momentum transfer and the corresponding projection parallel and perpendicular to the sample surface. Crystal axes are represented by black arrows. (c) A fitting example of the $dd$ excitations. The purple dotted line is an experimental spectrum, and the gray solid line represents the total fit. The inset depicts the energy splitting of 3$d$ orbitals.}
            \label{Fig1}
        \end{figure}
        
        \textit{Experimental details.}---Single-crystal \KCF compounds were prepared by the method described in Ref.\cite{Marshall2013}. A pristine sample with the surface normal (0 0 1) was selected and characterized by a lab-based Laue diffractrometer prior to the RIXS measurements. We confirm the sample has the type-$a$ orbital order structure (Fig. 5(a) in the Supplemental Material \cite{SM}). The RIXS experiments were conducted at the I21-RIXS beamline at Diamond Light Source, United Kingdom \cite{i21web}. The sample was mounted with the (1 1 0) plane lying in the scattering plane, as shown in Fig.\ref{Fig1} (b). The 2$\theta$ scattering angle was fixed at 146$^{\circ}$ throughout the experiment. RIXS measurements were also performed with the (1 0 0) plane lying in the scattering plane. The measuring temperature was kept at 16 K unless stated otherwise. We tuned the incident photon energy to the resonance of the Cu $L_3$ absorption peak [see Fig.\ref{Fig1} (a)] with either linear-horizontal ($\sigma$) or linear-vertical ($\pi$) polarizations for RIXS measurements. The total energy resolution is about 37 meV FWHM. RIXS signals were collected without polarization analysis. For all RIXS spectra, the elastic (zero-energy loss) peak positions were determined by the elastic scattering spectrum from carbon tape near the sample surface and then fine adjusted by the Gaussian fitted elastic peak position. All RIXS spectra are normalized by the integrated intensities from the high-energy region (0.5 eV $\sim$ 2 eV). The Miller indices in this study are defined by a pseudotetragonal unit cell with $a$ = $b \simeq$ 4.146\;\AA~and $c \simeq$ 3.92\;\AA. The momentum transfer $\textit{\textbf{q}}$ is defined in reciprocal lattice units (r.l.u.) as $\textit{\textbf{q}} = h\textit{\textbf{a}}^{*} + k \textit{\textbf{b}}^{*} + l \textit{\textbf{c}}^{*}$ where $\textit{\textbf{a}}^{*} = 2\pi/a$, etc.

        \textit{Results and discussion.}---Figure \ref{Fig1} (a) shows the Cu \textit{L}$_3$-edge x-ray absorption spectra (XAS) of \KCF excited by two linear polarizations. The main peak at 932.5eV corresponds to the $2p^53d^{10}$ final state, and the shoulder peak at about 935 eV stems from the $2p^53d^{10}\underline{L}$ state ($\underline{L}$ represents a hole at ligand-$F$ site) \cite{Nadai2001}. The comparable Cu \textit{L}$_3$-edge XAS intensity demonstrates the 3D character of the orbital ground state. A representative RIXS spectrum excited by $\pi$ polarized x-rays is shown in Fig. \ref{Fig1} (c), which comprises two regions: a high-energy $dd$ excitation that splits to four peaks, and a low-energy excitation region.
        
        \begin{figure}
            \centering
            \resizebox{8.6cm}{!}
            {\includegraphics{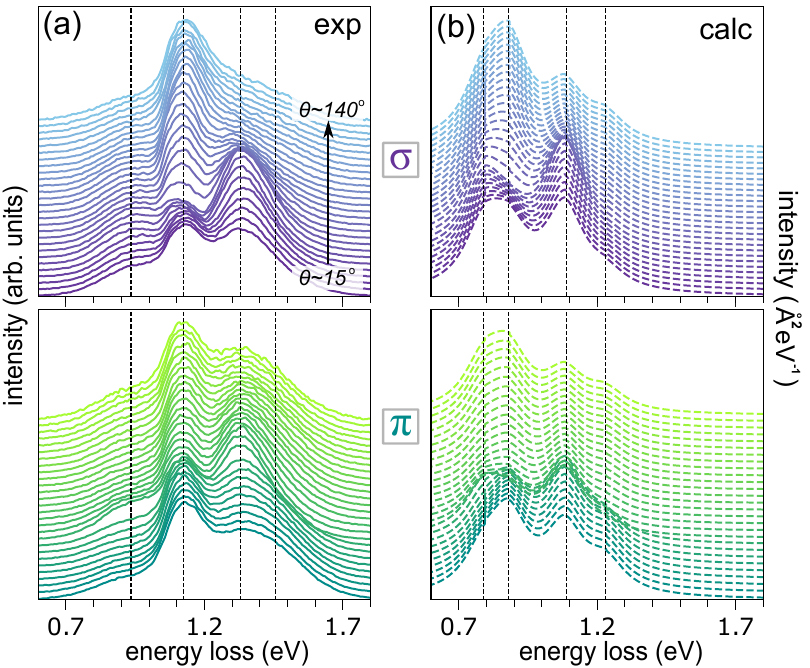}}
            \caption{Angular dependence of $dd$ excitations. (a) Experimental results from the $\sigma$- (purple) and $\pi$-polarized (green) incident x-rays. The vertical dashed lines depict the averaged peak values of fitted $dd$ excitations. (b) Calculated spectra from the MRCI+SOC. The vertical dashed lines show the calculated values of $dd$ excitations.}
            \label{Fig2}
        \end{figure}
        
        We first address the high-energy excitations. The $dd$ orbital excitations in \KCF, though having a comparable energy scale with respect to that of two-dimensional cuprates \cite{Sala2011}, possess different energy splitting owing to the low D$_{2h}$ crystal-field symmetry. We show the $3d$ orbitals splitting in the inset of Fig. \ref{Fig1}(c), where the ground state holds a $d_{x^2-z^2}$ hole orbital given the definition of the $xyz$ axes with respect to the crystal orientation \cite{SM}. The rotation of local distortion along each of three axes then induces the $d_{y^2-z^2}$ hole orbital at the next site, thus introducing a 3D long-range OO \cite{Hidaka1998}. As demonstrated in Fig. \ref{Fig1}(c), orbital excitations are resolved to four peaks labeled with orbital characters as sketched in the inset of Fig. \ref{Fig1}(c). We fitted the orbital excitations with a model comprising four Lorentzian functions convoluted by Gaussian energy resolution. Together we plot the fitted $dd$ peaks. The fitted energy positions are found to be comparable to optical and the Cu $K$-edge RIXS studies (see Table I in Ref. \cite{SM}) \cite{Deisenhofer2008, Ishii2011}.
        
        To further study the high-energy orbital excitations, we performed RIXS measurements by varying the incident $\theta$ angle from 15$^{\circ}$ to 140$^{\circ}$. The results are shown in Fig. \ref{Fig2}(a) with the top and the bottom figures from the $\sigma$ and the $\pi$ polarizations, respectively. Assisted by the fitting analysis, we reach the conclusion that all $dd$ excitations are nondispersive but exhibit a rich intensity variation as a function of $\theta$ (Fig. 6 in the Supplemental Material \cite{SM}). Similar to other cuprate compounds, the behavior in $dd$ excitations is known to be induced by the local ligand-field splitting \cite{Sala2011}.
        
        To understand better the $dd$ excitations, we carried out the \textit{ab initio} quantum chemistry calculations using the complete active space self-consistent field and multireference configuration interaction (MRCI) as implemented in the MOLPRO package \cite{Helgaker2000}. An embedded cluster consisting of a single CuF$_6$ octahedron (one Cu atom and six F atoms, with short and long bonding lengths in $ab$ plane) was considered in the calculations, using the crystallographic data as reported in Ref. \cite{Caciuffo2002}. In the MRCI treatment, the F 2$p$ and Cu 2$s$, 2$p$, 3$s$, 3$p$, 3$d$ electrons within the single CuF$_6$ unit were correlated. Details about the computed orbital excitation energies and the comparison with the experimental values are given in Ref. \cite{SM}. To account for the orbital ordering effect, the calculations were performed for both $d_{x^2-z^2}$ and $d_{y^2-z^2}$ hole orbitals. The latter was achieved by rotating the CuF$_6$ octahedron around $c/z$ axis by 90$^{\circ}$.

               \begin{figure}
            \centering
            \resizebox{8.6cm}{!}
            {\includegraphics{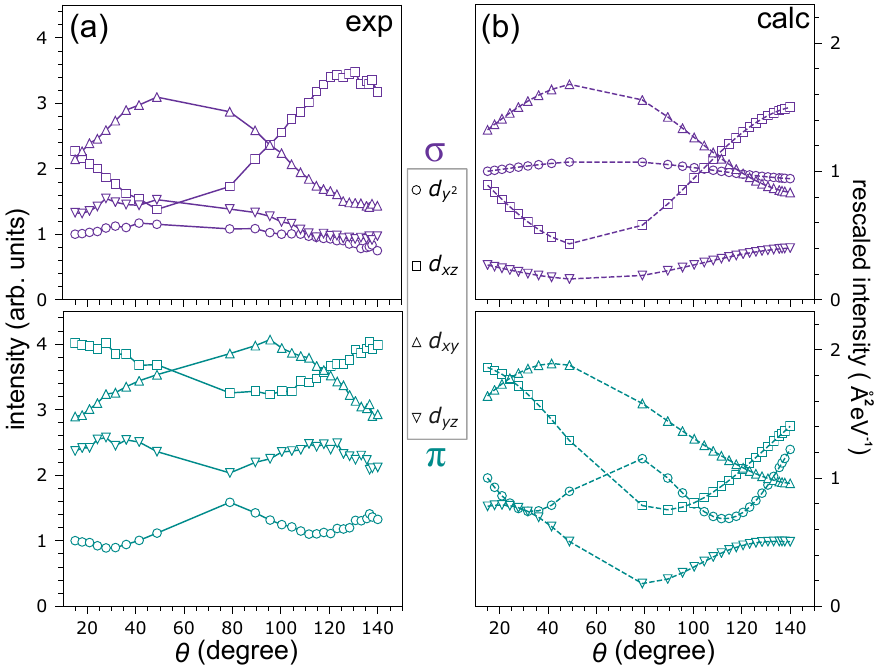}}
            \caption{Relative intensity variations as a function of $\theta$ angle for different $dd$ excitations. (a) is the experimental results; (b) is the calculated results.  The first data point of the $d_{y^2}$ orbital is normalized to a fixed intensity.}
            \label{Fig3}
        \end{figure}

        \begin{figure*}
            \centering
            {\includegraphics[width=\textwidth]{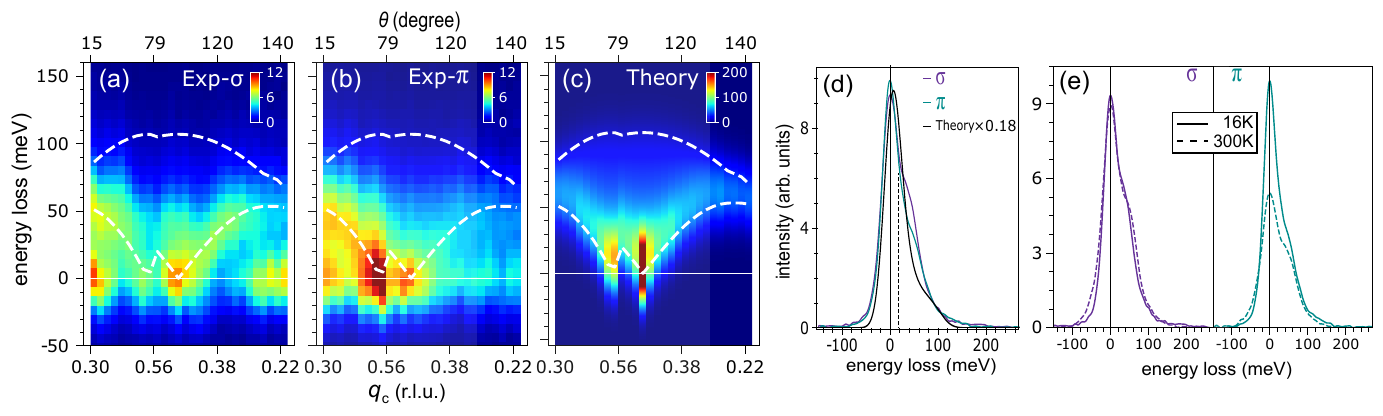}}
            \caption{Low-energy excitations of \KCF revealed by the Cu $L_3$-edge RIXS. (a) and (b) are color maps of the low-energy RIXS spectra from the $\sigma$ and $\pi$ polarizations, respectively. (c) is the Muller ansatz calculated results. The white dashed lines are the lower and upper boundaries of the two-spinon continuum. The thin white line is the zero energy reference. (d) The comparison of RIXS spectra at $q_c$=0.5 r.l.u. The dashed line marks the longitudinal magnetic mode observed in INS \cite{Lake2005}. (e) Temperature-dependent RIXS spectra at $q_c$=0.5 r.l.u. for different polarizations.}
            \label{Fig4}
        \end{figure*}

       Figure \ref{Fig2}(b) shows the averaged theoretical spectra as a function of $\theta$ based on MRCI plus spin-orbital coupling (SOC) approach. The results for both polarizations agree well with the experimental spectra. To make a more detailed comparison, we fitted all spectra and extracted the area of each orbital excitation. The intensities of each orbital excitation from the experimental and theoretical results are displayed in Fig. \ref{Fig3}(a), (b), respectively. The comparison shows fairly good agreement in terms of the trend of the angular-dependent intensity except for the $d_{yz}$ ($d_{xy}$) orbital in the $\sigma$ ($\pi$) polarization. The difference in the $d_{yz}$ orbital may be due to the uncertainty of determining the intensity because of the dominant $d_{xy}$ orbital. For the $d_{xy}$ orbital in the $\pi$ polarization, a repeated RIXS measurement shows consistent angular dependence. To cross check the theoretical results, we performed independent calculations using the single-ion model~\cite{Sala2011}. Interestingly, the results are consistent with the quantum chemistry calculations except for the $d_{xy}$ orbital in the $\pi$ polarization, which shows a maximum intensity around $\theta$ of 70$^{\circ}$ (Fig. 10 in the Supplemental Material~\cite{SM}). Therefore, the discrepancy likely stems from the simplification of the single cluster model. We note that the averaged results are similar to those derived from the single $d_{x^2-z^2}$ or $d_{y^2-z^2}$ hole orbital state (Fig. 7 in the Supplemental Material \cite{SM}). Remarkably, the experimental results acquired under the (1 0 0) geometry could only be supported by the averaged theoretical spectra than the spectra from either $d_{x^2-z^2}$ or $d_{y^2-z^2}$ single state (Figs. 8 and 9 in the Supplemental Material \cite{SM}). We therefore conclude that the high-energy $dd$ excitations are consistent with the local JT effect and the orbital ordering of the system. Altogether, the strongly localized nondispersive $dd$ excitations do not seem to support the existence of collective orbitons at such high-energy ranges \cite{Tanaka2004}. 

        We now turn to the discussion of the low-energy excitations to further explore the potential existence of the collective orbiton. In Fig. \ref{Fig4} (a), (b), we display the maps of the angular-dependent low-energy excitations probed by the linear $\sigma$ and $\pi$ polarizations, respectively. Both maps show excitations from zero loss energy up to about 100 meV, where a mode emanates from $\theta\simeq$ 100$^{\circ}$ and disperses to higher energy by approaching to either side of the $\theta$ range. Specifically, near a $\theta$ of 80$^{\circ}$, two maximal intensity points are present at the zero loss energy positions. In KCuF$_3$, the well-known dispersive modes are the two-spinon continuum reported by the inelastic neutron scattering (INS) because of the quasi-one-dimensional magnetic properties 
        ~\cite{Hutchings1969, Caciuffo2002, Satija1980}. Given the sensitivity of RIXS to magnetic excitations, we corroborate that the observed dispersive modes in RIXS are dominated by the two-spinon continuum.      
        
        To verify the assignment, we analyze more quantitatively the spin dynamics in \KCF using the Muller Ansatz \cite{Cloizeaux1962,Faddeev1981,Muller1981,Haldane1993}. As established in Ref.\cite{Lake2005_2}, the lower and upper boundaries of the two-spinon continuum can be expressed by the following sinusoidal functions:
        \begin{equation*}
            E_l(q_c) = \frac{\pi}{2} J_c |\textrm{sin}q_c| \quad \text{and} \quad E_u(q_c) = \pi J_c |\textrm{sin}\frac{q_c}{2}|,
            \label{Eq1}
        \end{equation*}
        where $q_c$ is the projected wave vector along the $c$ axis in the unit of r.l.u., and $J_{c}$ is the AFM superexchange interaction. As a good approximation, the magnetic dynamic structure factor at $T$=0 K can be expressed as \cite{Muller1981}
        \begin{equation*}
            S (E,q) = \frac{289.6}{\pi} \frac{H[E-E_l(q)] \times H[E_u(q)-E]}{\sqrt{E^2 - {E_l(q)}^2}}.
        \label{Eq2}
        \end{equation*}
        Here, $H[x]$ is the Heaviside step function. We evaluated this expression using $J_{c}=34$ meV which is based on the INS results \cite{Lake2005_2}. Before we plot the calculated spectra, we note that the photon momentum transfer along the sample $c$ axis passes through the AFM wave vector $q_c$ = 0.5 r.l.u. twice under the fixed RIXS scattering configuration \cite{SM}. Correspondingly, we show theoretical results as a function of $q_c$ in Fig. \ref{Fig4} (c) \cite{SM}. In Fig. \ref{Fig4} (a), (b), we plot the momentum transfer $q_c$ at the bottom axes and superimpose the two-spinon continuum lower and upper boundaries (white dashed lines) on top of the experimental results. Remarkably, the center of mass of the RIXS low-energy excitations matchesthe the lower limit of two-spinon dispersions extremely well. In particular, two maximal intensity spots near the zero loss energy position agree precisely with the theoretical results. 
        
        Comparing RIXS results to the INS data \cite{Lake2005_2}, some extra spectral weight seems to exist specifically near $q_c$ = 0.5 r.l.u.. We plot the corresponding line spectra in Fig. \ref{Fig4} (d) together with the theoretical result. The peak position of the mode appears at $\sim$40 (47) meV for the $\sigma$ ($\pi$) polarizations which is absent in the theory. The longitudinal magnetic mode, i.e., the signature of the 3D magnetic ordering, should in principle exist in RIXS spectra as the experiments were conducted below $T_N$ of 38 K \cite{Lake2005}. However, its center energy of $\sim$ 18.5 meV is below the RIXS energy resolution. On the other hand, the mode seems to be comparable to the optical phonon observed by Raman \cite{Lee2012} and IXS \cite{Tanaka2004}. To further explore the origin of the mode, we performed temperature-dependent measurements at $q_c$=0.5 r.l.u. The data are shown in Fig. \ref{Fig4} (e). The persistence of the peak up to room temperature in both polarizations confirms the phonon-like origin of this low-energy mode at $q_c$=0.5 r.l.u.
        
        We now address the relevance of the observed low-energy excitations to the expected dispersive orbiton in KCuF$_3$. It is understood that the spin-orbital exchange of purely electronic origin should not be considered the mere mechanism that is responsible for the orbiton dispersion. As the OO is mostly driven by the JT mechanism whose Hamiltonian has an identical form to the orbital part of the superexchange interaction~\cite{Kugel1982, Okamoto2002, Brink2004, Nasu2013}, the orbiton dispersion should mainly result from the JT effect. This is consistent with what was suggested in Ref. \cite{Lee2012}, which is that the orbital-only interaction, which comprises the JT and the on-ligand interaction, is about 600 K ($\sim$ 52 meV). By taking into account also the effective spin-orbital exchange of electronic origin ($\sim$ 3 meV), we estimate approximately the total orbital superexchange ($J_{\rm OO}$) to be about 55 meV along the $c$ direction, which could result in an orbiton bandwidth of $\sim$ 110 meV. Such a large bandwidth would yield an obvious dispersive orbiton given the experimental energy resolution of 37 meV. 

Interestingly, we found an excellent agreement between the low-energy RIXS spectra and the Muller ansatz calculations, which indicates
that the orbiton dispersion must be well below the dispersive two-spinon continuum. It is thus puzzling to understand why the orbital superexchange ($J_{\rm OO}$ $\sim$ 55 meV) induced JT coupling does not lead to a sizeable and observable orbiton dispersion. While we leave a detailed answer to this question
for a future work, we here suggest the following explanation. Apart from the cooperative, global JT effect and ordering, the
local JT effect, i.e., the JT coupling between the orbital degrees of freedom and the local lattice vibrations, is of prime importance \cite{Nasu2013, Brink2001}. As discussed in Ref. \cite{Nasu2013}, the latter effect, which should be present in any
JT active system, can lead to a strong dressing of the orbitons with local vibrational modes and
may cause a complete smearing out of the orbiton dispersion, cf. Fig. 5(b) of ~\cite{Nasu2013}.
Since the JT coupling is inherently strong in KCuF$_3$, we believe this scenario
explains the effective disappearance of the orbiton dispersion in this system. Our work is significant in recognizing the importance of the dressing effect of the local JT distortion on the collective orbiton in \KCF and many other orbital-ordered systems with strong JT effects.

        \textit{Conclusion.}---In summary, we performed high-resolution RIXS experiments on the orbitally ordered \KCF. The high-energy excitations are found to stem from localized $dd$
orbital excitations, consistent with the {\it ab initio} calculation based on a single cluster. The low-energy dispersive excitations are dominated by the two-spinon continuum via the comparison to Mueller ansatz calculations. This indicates that the relevant energy bandwidth of the orbitons may be much lower than the energy resolution of RIXS experiments. We suggest that the main reason for the lack of the onset of an orbiton with a dispersion above the resolution threshold lies in the possibly strong local JT effect, which may lead to
the dressing of the orbiton with the local vibrational modes and thus to the suppression of the orbiton dispersion. 
 
        We thank E. Pavarini for fruitful discussions. We also thank Nikolay Bogdanov for the assistance on the theoretical models. J.L. acknowledges Diamond Light Source (United Kingdom) and the Institute of Physics in Chinese Academy of Sciences (China) for providing funding Grant No. 112111KYSB20170059 for the joint Doctoral Training under the contract STU0171, and also the financial support from China Scholarship Council. L.X. thanks U. Nitzsche for technical assistance. J.S.Z. acknowledges the support by NSF Grant No. DMR-1905598 in the United States. K.W. acknowledges support by Narodowe Centrum Nauki (NCN) Project No. 2016/22/E/ST3/00560. J. vdB acknowledges financial support through the Deutsche Forschungsgemeinschaft (DFG, German Research Foundation), SFB 1143 project A5, and through the W{\"u}rzburg-Dresden Cluster of Excellence on Complexity and Topology in Quantum Matter--ct.qmat (EXC 2147, project-id 39085490). H.D. acknowledges support by the National Natural Science Foundation of China (No. 11888101) and the Ministry of Science and Technology of China (2016YFA0401000). All data were taken at the I21 RIXS beam line of Diamond Light Source (United Kingdom) using the RIXS spectrometer designed, built and owned by Diamond Light Source. We acknowledge Diamond Light Source for providing the science commissioning time on Beamline I21. We acknowledge Thomas Rice for the technical support throughout the beam time. We would also like to thank the Materials Characterization Laboratory team for help on the Laue instrument in the Materials Characterization Laboratory at the ISIS Neutron and Muon Source.

        J.L. and L.X. contributed equally to this work.

\pagebreak

 \section{Supplementary Material}

\maketitle
 \section{I. Sample Information}
\begin{figure}
            \centering
            \resizebox{8.6cm}{!}
            {\includegraphics{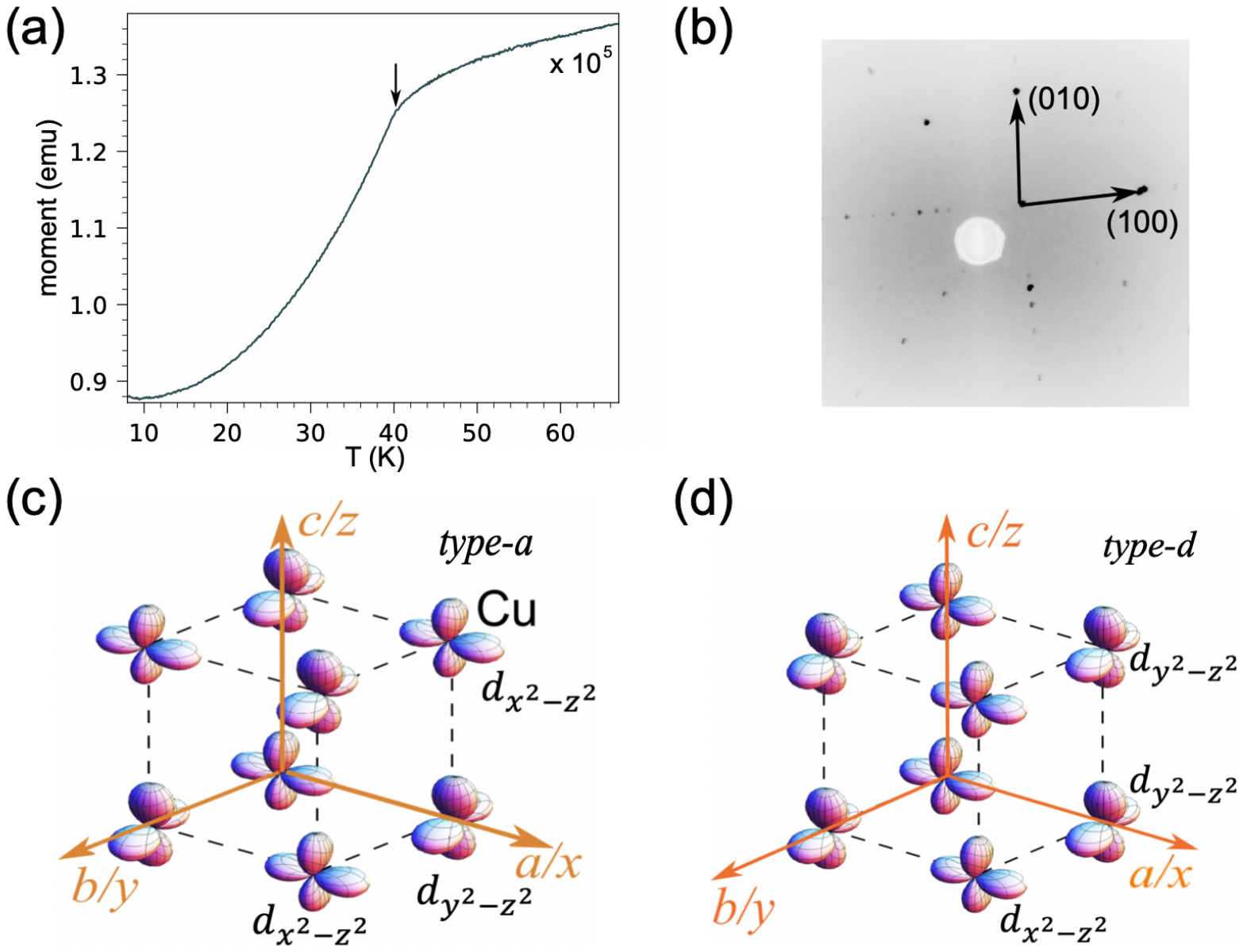}}
            \caption{ (a) Magnetic susceptibility of measured sample. The black arrow indicates the transition temperature is $\sim$39 K . (b) The sample's Laue pattern with the in-plane orientations marked by the black arrows. (c) and (d) depicts the type -$a$ and -$d$ orbital order structure respectively.}
            \label{SM1}
        \end{figure}

        We used Physical Property Measurement System to measure the magnetic susceptibility of the same crystal used by RIXS experiments. Fig.\ref{SM1}(a) shows the measured magnetic susceptibility. The $T_N$ is found to be around 39 K. To pre-align the sample prior the RIXS measurements, we utilized the lab-based X-ray diffractometer to obtain the crystal's diffraction pattern. As shown in Fig.\ref{SM1}(b), the sample surface normal was confirmed to be (001) and the in-plane directions were also characterized. The angular offsets of the sample were corrected prior to the RIXS experiments. In KCuF$_3$, the alternation of $d_{x^2-z^2}$ and $d_{y^2-z^2}$ orbitals along three crystalline directions defines two types of orbital ordering structure [see Fig.\ref{SM1}(c) and (d)]: the type--$a$ where the orbitals between two adjacent CuF planes are rotated by 90$^\circ$ and the type--$d$ where they remain identical vertically \cite{Tsukuda1972}. The spin orders in the same way for both types of structure according to the $Goodenough-Kanamori$ rules, but exhibits different transition temperature ($T_N\sim$38  K for type-$a$ and $T_N\sim$22 K for type-$d$) \cite{Hutchings1969}. The magnetic susceptibility shown in Fig.\ref{SM1}(a) confirms that our sample has the type-$a$ orbital order structure.

    \section{II. $dd$ excitations}
        \begin{table*}[b]
            \caption{
                Comparison of energy splittings for different orbitals between fitted results and theoretical calculations. The Jahn-Teller distortions occur within the $xy$ plane. Each MRCI+SOC value stands for a spin-orbit Kramers doublet. The calculated energies for Cu$^{2+}$ $3d^{9}$ and $2p^{5}3d^{10}$ states are also given. All the energies are in the unit of eV.
            }
            \begin{ruledtabular}
            \begin{tabular}{c c c c c c c}
                 Hole orbital  &    CASSCF  &   MRCI  &   MRCI+SOC   &   Exp.  &  Ref.\cite{Deisenhofer2008}  & Ref.\cite{Ishii2011}\\
                \hline
                $d_{x^2 - z^2}$.    &  0.00          & 0.00      & 0.00       & 0.00  & 0.00 & 0.00 \\
                $d_{y^2}$           &  0.76          & 0.82      & 0.79       & 0.91  & 1.02 & $\sim$1\\
                $d_{xz}$            &  0.82          & 0.89      & 0.88      & 1.12  & 1.15  & *\\
                $d_{xy}$            &  0.98          & 1.05      & 1.09       & 1.32  & 1.37 & $\sim$1.3\\
                $d_{yz}$            &  1.08          & 1.16      & 1.23       &  1.45 & 1.46 & *\\
                $p^{3}_{3/2}$  	     &  940.50/940.54    &  939.18/939.21  & 932.46/932.52	     & Exp.934.8	& --- & --- \\ 
                $p^{1}_{1/2}$   	     &   940.60                &  939.28               & 952.74	     & --- & --- & --- \\              
            \end{tabular}
            \label{Cu_d9}  \footnotetext[1]{CASSCF: complete active space self-consistent field, MRCI: multireference configuration-interaction, SOC: spin-orbital coupling.}
            \footnotetext[2]{Limited by the energy resolution, the Cu $K$-edge RIXS study in Ref.\cite{Ishii2011} can only resolve two peaks with one coming from the transition between two $e_g$ orbitals and the other one from the transition between the $e_g$ and $t_{2g}$.}
            \end{ruledtabular}
        \end{table*} 

        \begin{figure}
            \centering
            \resizebox{8.6cm}{!}
            {\includegraphics{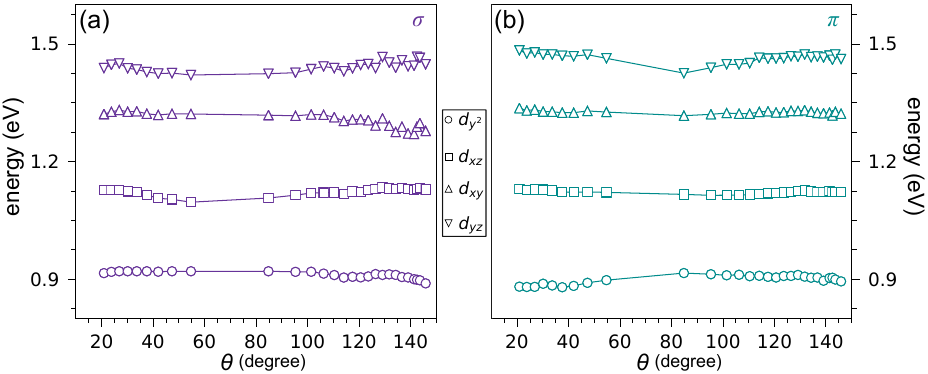}}
            \caption{Fitting energies for different orbitals as a function of the $\theta$-angle. (a) and (b) are from the $\sigma$- and $\pi$- polarization respectively.}
            \label{SM4}
        \end{figure}
        
        As described in the main text, the $dd$ excitations were fitted with a model composing four Lorentizian profiles convoluted with the instrumental energy resolution. In Fig.\ref{SM4}, we display the fitted energies of different orbitals as a function of $\theta$-angle. All $dd$ excitations energy positions are invariable within the error bar. The mean values of fitted peak energy of each orbital are summarized in Table \ref{Cu_d9}, together with the calculated results from various models [see section IV]. It is interesting to note that both CASSCF and MRCI tend to slightly underestimate the measured orbital energy by 0.2 -- 0.3 eV. Comparing to previous optical and the Cu K-edge RIXS studies, our results show good consistency \cite{Deisenhofer2008, Ishii2011}.
        
\begin{figure}
            \centering
            \resizebox{8.6cm}{!}
            {\includegraphics{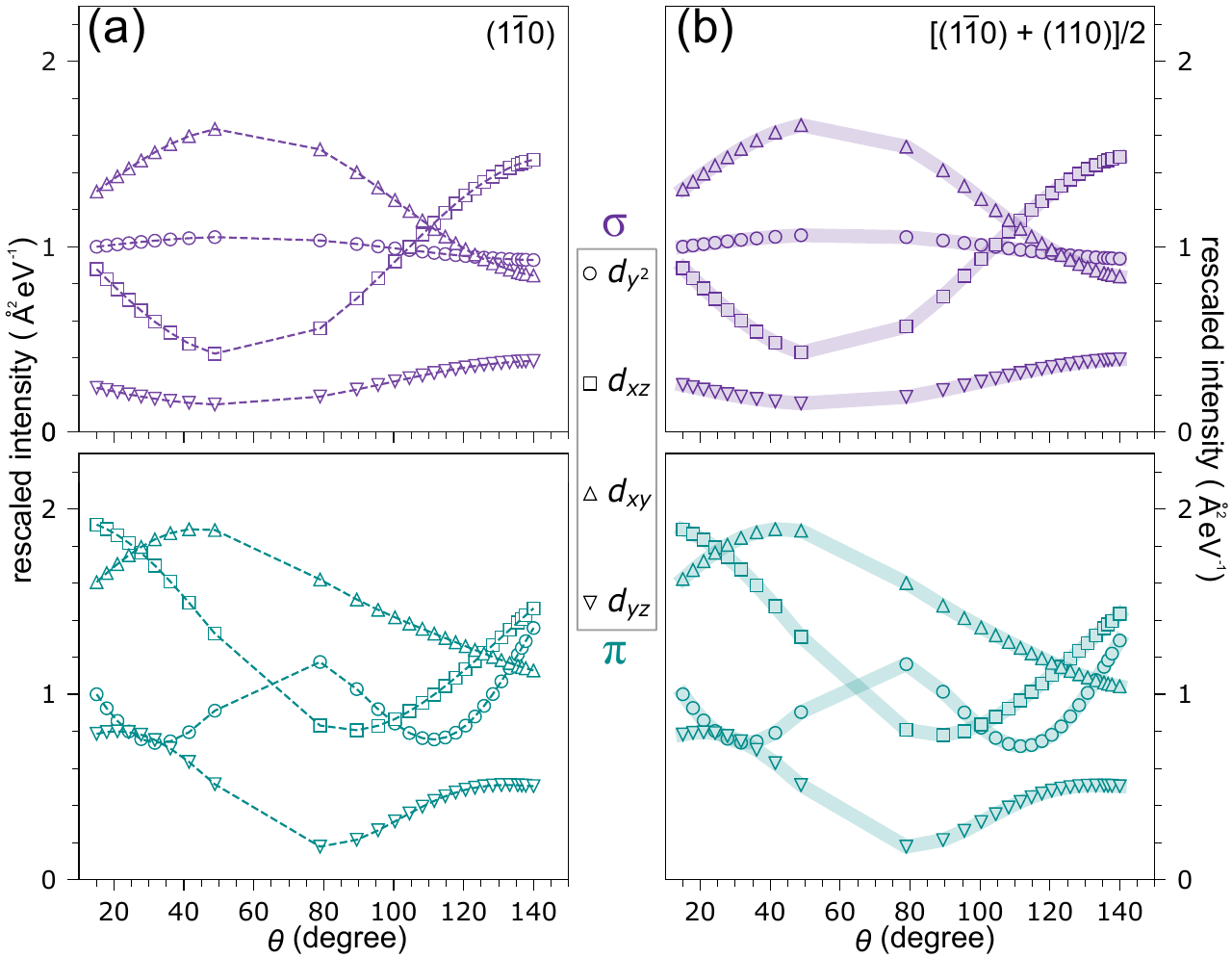}}
            \caption{ Relative intensity variations as a function of the $\theta$ angle for different $dd$ excitations. (a) is the calculated results from the $d_{y^2-z^2}$ ground state and (b) is the average of calculations from the $d_{x^2-z^2}$ and $d_{y^2-z^2}$ ground states.}
            \label{SM5}
        \end{figure}

        To account for the long range orbital order in QC calculation, we rotate the single CuF$_6$ octahedron along $z$-axis [see Fig.\ref{SM2}] by 90$^{\circ}$ so that the orbital of ground state in RIXS process becomes $d_{y^2-z^2}$ and the scattering plane changes to the (1-10). The intensity variations for each orbital are summarized in Fig.\ref{SM5}. It is clear that the average of $d_{x^2-z^2}$ and $d_{y^2-z^2}$ is very similar to the individual ones obtained from the pure $d_{x^2-z^2}$ or $d_{y^2-z^2}$ orbital ground state. 
        
\begin{figure*}
            \centering
            {\includegraphics[width=\textwidth]{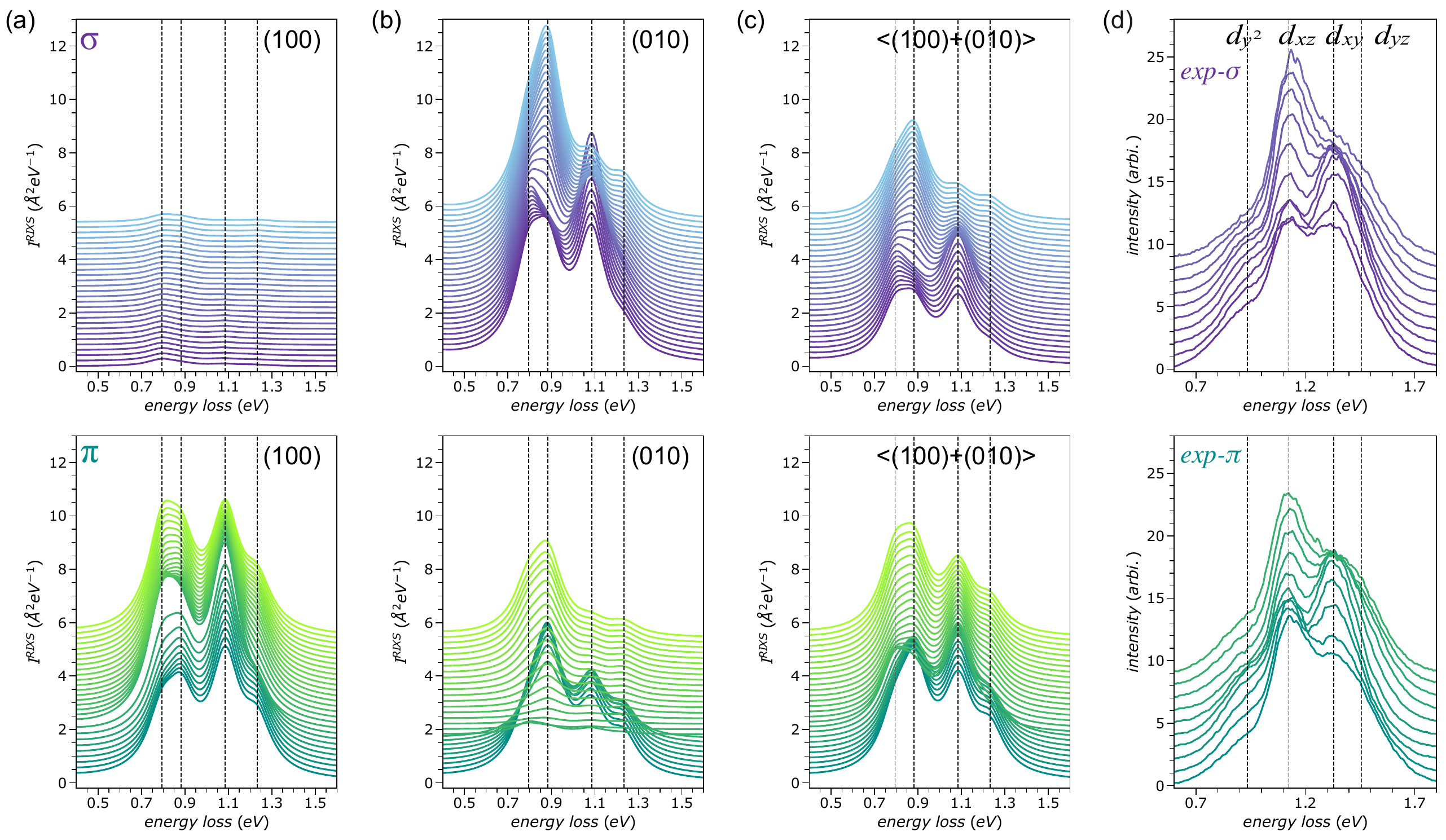}}
            \caption{The $dd$ excitations with the (100) plane lying in the scattering plane. $QC$ theoretical RIXS spectra: (a) of $d_{x^2-z^2}$ hole orbital, (b) of $d_{y^2-z^2}$ hole orbital, (c) of the averaged hole orbital states; (d) the experimental results. }
            \label{Fig R3}
        \end{figure*}

\begin{figure*}
            \centering
            {\includegraphics[width=\textwidth]{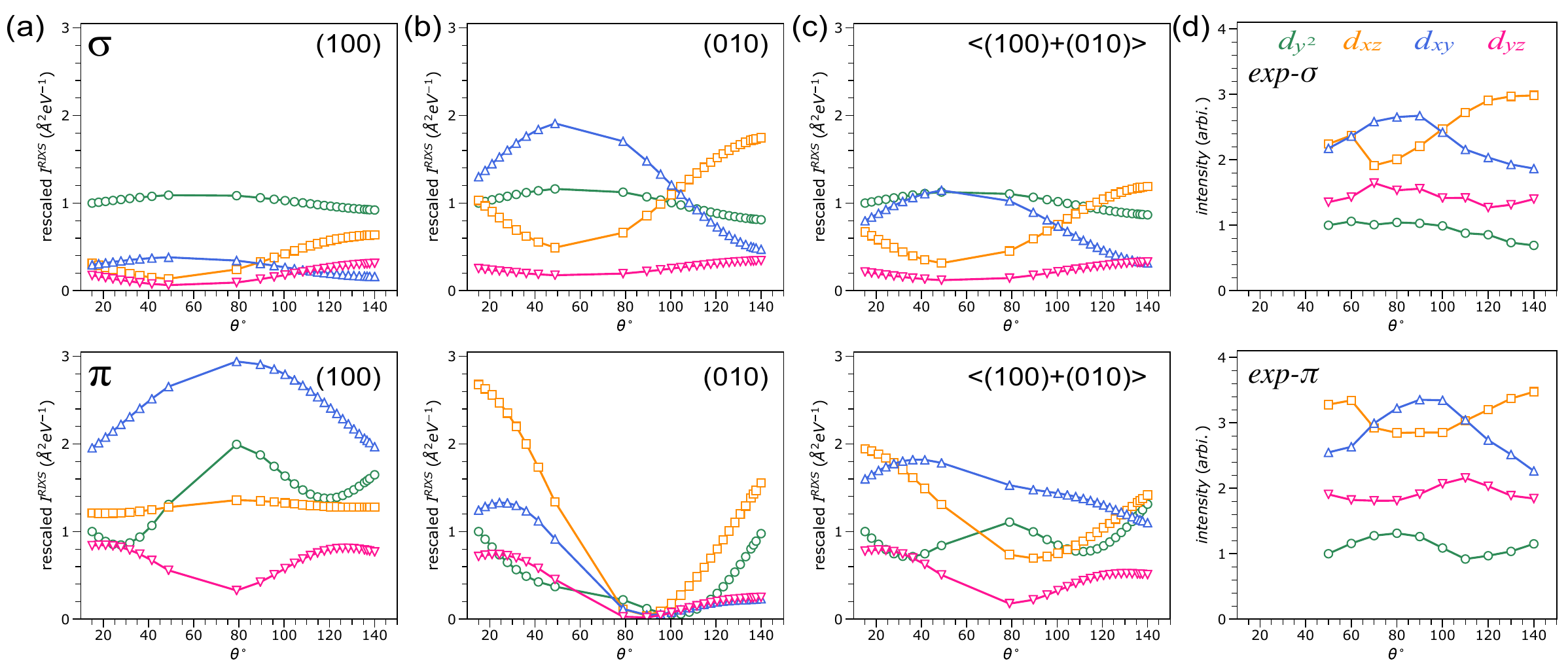}}
            \caption{The extracted dd intensities from FIG.\ref{Fig R3}, as a function of $\theta$ angle.}
            \label{Fig R4}
        \end{figure*}

        The RIXS results acquired with the (100) plane lying in the scattering plane were presented in Fig.\ref{Fig R3}. Together were shown the theoretical QC calculated spectra for the individual orbital ground states $d_{x^2-z^2}$,  $d_{y^2-z^2}$, as well as the averaged ones. We extracted the fitted peak intensities and summarized in Fig.\ref{Fig R4}. The experimental results are consistent with the averaged theoretical spectra instead of individual ones. 
        
     \section{III. Single-ion calculation}
\begin{figure}
            \centering
            \resizebox{8.6cm}{!}
            {\includegraphics{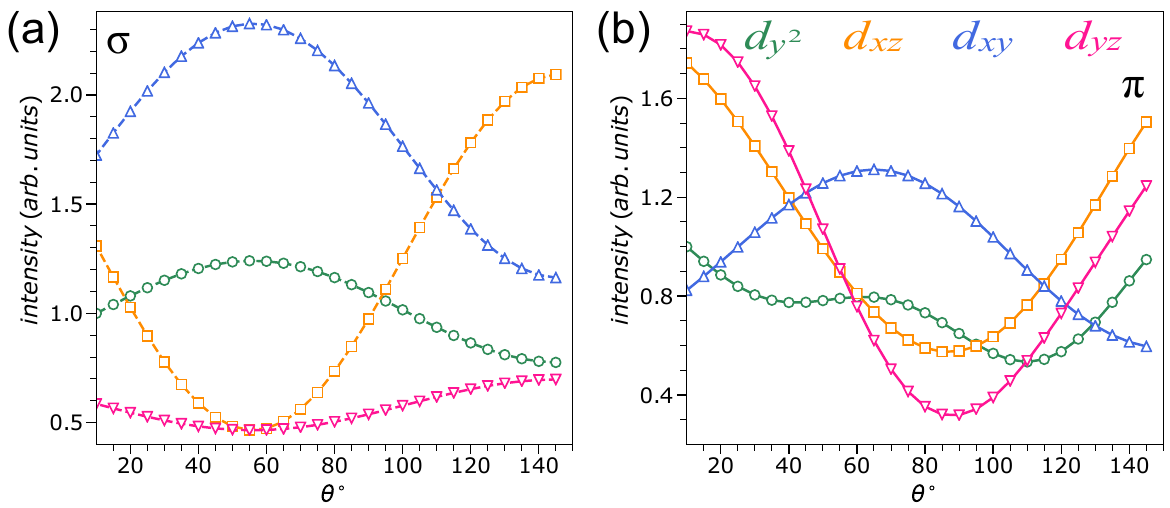}}
            \caption{The intensities of each $dd$ orbital excitations computed using the single-ion model.}
            \label{SM_singleion}
        \end{figure}

        The single-ion model was used to calculate the dd orbital excitation for cross-checking the intensity variation as a function of incident angle~\cite{Sala2011}. Results are shown in Fig.\ref{SM_singleion}. The results are
consistent to the QC calculations except the $d_{xy}$ orbital in $\pi$ polarization which shows a maximum around $\theta$ of 70$^{\circ}$. 

    \section{IV. Quantum Chemistry Calculation}
\begin{figure}
            \centering
            \resizebox{8.6cm}{!}
            {\includegraphics{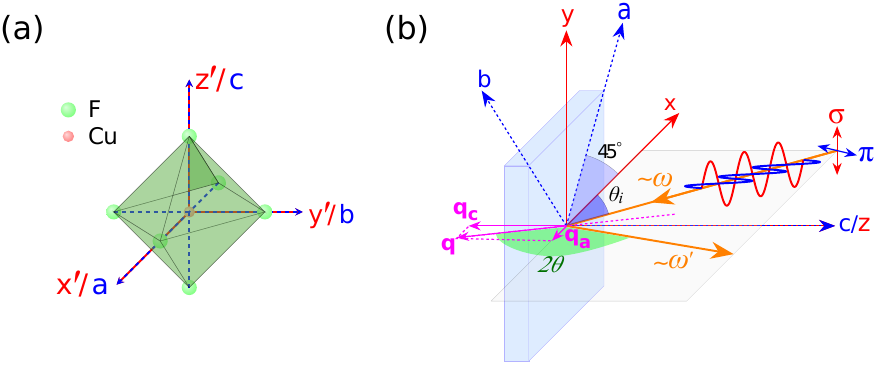}}
            \caption{(a)The CuF$_6$ cluster used in quantum chemistry calculations. (b) The RIXS scattering geometry.}
            \label{SM2}
        \end{figure}

		QC calculations were carried out on an embedded cluster consisting of one CuF$_{6}$ octahedron [see Fig.\ref{SM2}(a)]. Crystallographic data as reported in Ref.\cite{Buttner1990} were employed to construct the cluster and the embedding potential. The latter corresponds to a finite array of point charges fitted to reproduce the crystal Madelung field in the cluster region \cite{Klintenberg2000}. To describe the readjustment of the charge distribution in the vicinity of an excited electron in the RIXS processes, we use a nonorthogonal configuration-interaction type of approach. This approach is based on a separate self-consistent filed (SCF) optimizations for the many-body wave functions describing the reference $3d^{9}$ and core-hole $2p^{5}3d^{10}$ configurations, the same methodology as discussed in Ref. \cite{Bogdanov2017}. For the Cu ion, we used all-electron triple-zeta Douglas-Kroll basis sets with diffuse functions and weighted core-valence sets (aug-cc-pwCVTZ-DK) \cite{Balabanov2005} to capture core valence correlation effects. For the six F ligands, all-electron double-zeta Douglas-Kroll basis sets with diffuse functions (aug-cc-pVDZ-DK) \cite{Dunning1989} were used. 
		
		\textbf{Cu$^{2+}$ $3d^{9}$ valence-excited states.} To access the on-site Cu$^{2+}$ 3$d^9$ valence-excited states, we used an active space of  five $3d$ orbitals ($t_{2g}$ and $e_{g}$) and nine electrons in the complete active space self-consistent field (CASSCF) \cite{Helgaker2000} calculations. The SCF optimization was carried out for an average of all five doublet states associated with this manifold. The Pipek-Mezey localization module\cite{Pipek1989} available in {\sc molpro} was employed for localizing the Cu and F orbitals. On top of the CASSCF reference, the multireference configuration-interaction (MRCI) \cite{Knowles1992,Werner1988} expansion additionally includes single and double excitations from  the Cu $2s, 2p, 3s, 3p, 3d$ shells and the $2p$ orbitals of the F ligands. All five valence-excited doublet states entered the spin-orbit calculations, both at the CASSCF and MRCI levels.

		\textbf{Cu$^{2+}$ $2p^{5}3d^{10}$ core-hole states.} For the intermediate $2p^{5}3d^{10}$ core-hole states,  the CASSCF treatment was performed in terms of fifteen electrons and five Cu $3d$ orbitals plus three Cu $2p$ orbitals while freezing these Cu $2p$  orbitals (with the occupation restriction of having maximum five electrons) in the active space. The SCF optimization was carried out for an average of three doublet states associated with the $2p^{5}3d^{10}$ configuration. Appropriate changes were operated in the subsequent MRCI treatment such that the Cu $2p$ orbitals have at most five $2p$ electrons for internal and semi-internal substitutions. The Cu $2s, 2p, 3s, 3p, 3d$ shells and F $2p$ shells were correlated. The MRCI was performed as three-root calculation; these three core-hole states further entered the spin-orbit calculations.
		
		\textbf{Dipole transition matrix elements.} The individual SCF optimizations of the valence-excited and core-hole states leads to sets of nonorthogonal orbitals. The MRCI dipole transition matrix elements between wave-functions expressed in terms of such nonorthogonal orbitals for the $3d^{9}$ and $2p^{5}d^{10}$ groups of states (i.e., five valence-excited and three core-hole states) were derived according to the procedure described in Ref. \cite{Mitrushchenkov2007}. 
		
		For the analysis of the quantum chemistry wave functions, we use a local coordinate frame \{$x^{\prime}$, $y^{\prime}$, $z^{\prime}$\} with the $x^{\prime}$, $y^{\prime}$ and $z^{\prime}$ as indicated in Fig.\ref{SM2}(a). The transformation to the coordinate frame of experimental scattering geometry \{$x$,$y$,$z$\}  is straightforward, by $45^{\circ}$ clockwise rotation around the $c/z$ axis as shown in Fig.\ref{SM2}(b). The rotation of the $\sigma$, $\pi$ and $\pi^{\prime}$ (introduced as outgoing $\pi$ polarization) vectors as function of the angle $\theta$ is described by geometrical relations:
		
                \begin{equation}
                    \begin{aligned}
                    \label{dm_rotation}
                    \vec{D}_{\sigma}  &  = \vec{D}_{y}, \\
                    \vec{D}_{\pi}   & = \vec{D}_{x}\sin\theta + \vec{D}_{z} \cos\theta, \\
                    \vec{D}_{\pi^{\prime}}   & = \vec{D}_{x}\sin(\theta + \alpha) + \vec{D}_{z} \cos(\theta+\alpha),
                    \end{aligned}
                \end{equation}
                where $\vec{D}^{kl}_{\epsilon}$=$ \braket{ \Psi^{k}_{\text{fs}} | \vec{\epsilon}\cdot \vec{R} | \Psi^{l}_{\text{c}^{\star} }}$ stands for dipole transition matrix elements.
                
		The RIXS double differential cross section \cite{Veenendaal2015, Bogdanov2017} can be computed by summing over the outgoing polarization directions:
		    \begin{widetext}
                \begin{align*}
                 &   I^{\!R\!I\!X\!S}(\hbar\omega, \hbar\omega^{\prime}, \epsilon, \theta)     =\frac{ \text{d}^2 \sigma^{\text{RIXS}}_{\vec{k}, \epsilon }  }
                  { \text{d} {\Omega^{\prime}} \text{d} \hbar \omega^{\prime}}  =  \\
                 &  \frac{\alpha^2 \hbar^2}{e^4c^2}\omega {\omega^{\prime}}^3  
                   \sum_{\epsilon^{\prime}}  \sum_{j} \frac{1}{\textrm{g}_{\text{gs}}}\sum_{k}  
                      \abs{
                     \sum_{l} 
                    \frac{ \braket{ \Psi^{k}_{\text{fs}} | \vec{D}_{\epsilon^{\prime}}  | \Psi^{l}_{\text{c}^{\star}}  } 
                          \braket{ \Psi^{l}_{\text{c}^{\star}} | \vec{D}_{\epsilon} | \Psi^{j}_{\text{gs}}  }}       
                          { E^{j}_{\text{gs}} + \hbar \omega  - E^{l}_{\text{c}^{\star}} + (i\Gamma^{}_{\text{c}^{\star}
                          }\!/2) } }^{2}  \notag \times \frac{ \Gamma^{}_{\text{fs}}/2\pi}{ (E^{j}_{\text{gs}} + \hbar \omega - E^{k}_{\text{fs}} - \hbar \omega^{\prime})^2 + {(\Gamma^{}_{\text{fs}})}^{2}\!/4 } \numberthis. 
                \label{RIXS_sum}	                      
                \end{align*}
            \end{widetext}

		In Eq.\ref{RIXS_sum}, $\alpha$ is the fine structure constant, $\hbar\omega$ is the energy of the incoming photons, $\hbar\omega^{\prime}$ is the energy of outgoing photons, $\epsilon$ and $\epsilon^{\prime}$ are polarizations of the incoming and outgoing photons, respectively.
        For $I^{\!R\!I\!X\!S}$, the summations take into account all core-hole (intermediate) states and $2p^63d^9$ (final) states and the possible degeneracy of the ground state $\textrm{g}_{\text{gs}}$. The lifetimes for the core-excited and valence-excited states are $\Gamma^{}_{\text{c}^{\star}}$ and $\Gamma^{}_{\text{fs}}$, respectively. The $\Gamma^{}_{\text{c}^{\star}}$ and $\Gamma^{}_{\text{fs}}$ are here set to 1 eV and 0.16 eV, respectively, typical values for Cu $L$-edge RIXS \cite{Sala2011, Luuk2011}.
        
    \section{V. Spin dynamics}
\begin{figure}
                \centering
                \resizebox{8.6cm}{!}
                {\includegraphics{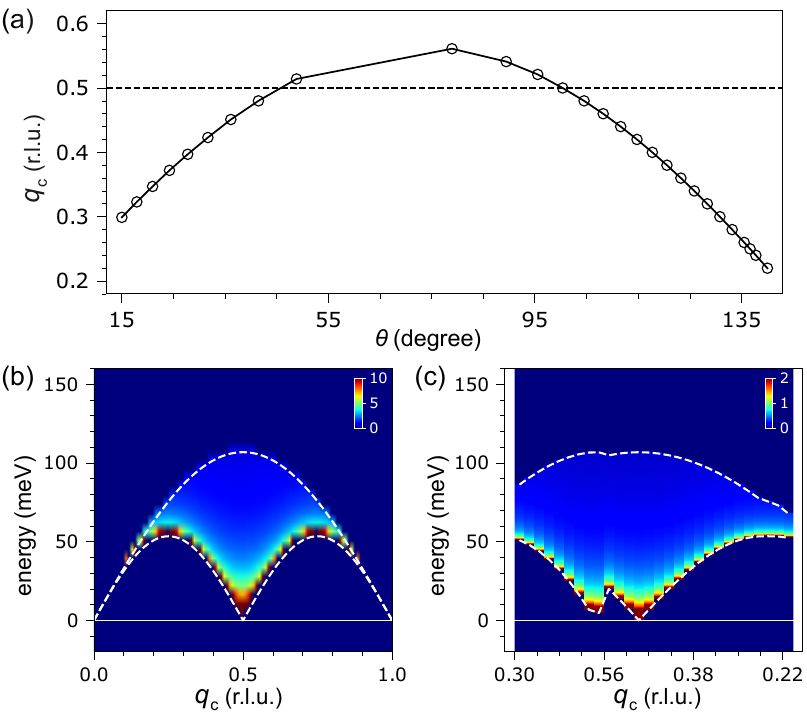}}
                \caption{Two-spinon continuum. (a) Relation between $q_c$ $r.l.u.$ and rotating angle $\theta$. The dashed line indicates where the $q_c$=0.5 $r.l.u.$ is. (b) Result of the standard projection along the $c$-axis  direction. (c) The specific projection along the $c$-axis constrained by the experimental geometry with a fixed scattering angle.}
                \label{SM3}
            \end{figure}

        Assisted by the underlying orbital order, the spin degree of freedom experiences a strong AFM coupling ($J_c$=34 meV) along the $c$-axis, confining KCuF$_3$ to a one-dimensional AFM Heisenberg system \cite{Satija1980}. Due to large quantum fluctuation, the ground state of such system is disordered with decaying correlations \cite{Cloizeaux1962} and the the corresponding excitation is the two-spinon continuum \cite{Faddeev1981} which carries a spin quantum number of $\frac{1}{2}$ and always appears in pairs \cite{Haldane1993}. As a good approximation, the magnetic dynamic structure factor (MDS) at $T$=0 K can be expressed by the Muller ansatz \cite{Muller1981} [see the discussion in main text]. In Fig.\ref{SM3}, we simulated the Muller ansatz for two different geometry: one is the standard projection along the $c$-axis direction, and the other one is the specific projection along the $c$-axis constrained by the experimental geometry with a fixed scattering angle. 
        
\end{document}